\documentclass[journal]{IEEEtran}

\ifCLASSINFOpdf
\else
\fi

\usepackage{array}
\usepackage{times}
\usepackage{epsfig}
\usepackage{graphicx}
\usepackage{amsmath}
\usepackage{amssymb}
\usepackage{graphicx}
\usepackage[utf8]{inputenc}
\usepackage{textcomp}
\usepackage{cite}
\usepackage{enumerate}
\usepackage{cases}
\usepackage{multirow}
\usepackage{verbatim}
\usepackage{amssymb}
\usepackage{CJK}
\usepackage{algorithm}
\usepackage{algorithmicx}
\usepackage{algpseudocode}
\usepackage{stfloats}
\usepackage{color}
\usepackage[dvipsnames]{xcolor}
\usepackage{bm}
\usepackage{booktabs}
\usepackage[colorlinks,linkcolor=red]{hyperref}
\usepackage{makecell}
\usepackage{tabularx}

\begin{document}

\title{Estimation of Specific Gravity of Potato Tubers Using Dielectric Properties}


\author{Taorui Chen, Yuki Gao, \IEEEmembership{Graduate Student Member,~IEEE,} Yi Wang, and Hai-Han Sun, \IEEEmembership{Senior Member,~IEEE}

}

\markboth{}%
{Shell \MakeLowercase{\textit{et al.}}: Bare Demo of IEEEtran.cls for IEEE Journals}

\maketitle

\begin{abstract}
Potatoes are an economically important crop, and their quality is closely related to the starch content, which is typically inferred from specific gravity (SG). Although microwave sensing technologies have been increasingly developed for underground potato detection and quality assessment in recent years, no accurate model has yet been established to link the dielectric properties of potatoes with their key agronomic traits. To address this gap, we developed a model for estimating potato tubers' SG based on their dielectric constant. To construct and validate the model, we conducted SG measurements and dielectric spectroscopy measurements in the frequency range of 0.3 GHz to 3.0 GHz on 250 potatoes of five different types (red, russet, yellow, purple, and chipping potatoes, with 50 samples per type). Out of the 250 data sets, 200 data sets were used for model development, and 50 data sets were used for model validation. A linear regression model was used to summarize the relationship between SG and dielectric constant, where the regression coefficients are expressed as fourth-order polynomial functions of frequency. Experimental results on the 50 validation data sets show that the model achieves high estimation accuracy with mean absolute errors (MAE) less than \(4.8\times10^{-3}\) and mean absolute percentage errors (MAPE) less than 0.45\%. The study of the dielectric properties of potatoes, along with the derived SG estimation model, provides a foundation for the future development of microwave sensing technologies for agronomic trait assessment in the potato production and processing industries. All measured data will be made publicly available upon acceptance of the paper.
\end{abstract}

\begin{IEEEkeywords}
Agricultural datasets, dielectric properties, linear regression model, potato tubers, specific gravity
\end{IEEEkeywords}

\IEEEpeerreviewmaketitle
\section{Introduction}

\IEEEPARstart {P}{\lowercase{o}}tato is the fourth most consumed staple food worldwide and plays a critical role in ensuring global food security~\cite{Intro1}. One of the key traits for potato quality assessment, usage classification, and harvest timing estimation is starch content, which is typically inferred from the specific gravity (SG) of the potato ~\cite{SG1, SG2, SG3}. Traditionally, measuring the SG of potatoes during the growing season requires harvesting samples from a small area, conducting SG measurements, and then using the results to estimate field-wide conditions. This method is destructive, time-consuming, and incapable of providing accurate field-scale assessments due to variability in potato growth across fields~\cite{rgrow}. To overcome these limitations, microwave sensing methods~\cite{mics1, mics2, mics3, mics4} have been increasingly explored for non-destructive, on-site potato measurements. Microwave sensing techniques measure and derive the dielectric properties of targets~\cite{rmwd1,rmwd2,rmwd3}, which are governed by their internal compositions. For potatoes, dielectric properties are primarily influenced by dry matter and moisture content, and the ratio of dry matter to moisture content determines the specific gravity (SG). Thus, dielectric properties have the potential to directly infer SG. However, to date, no accurate model has been established between potatoes’ dielectric properties and SG. Developing a robust model to bridge this gap would significantly advance the development of non-destructive microwave sensing technologies for potato trait characterization in the food production and processing industries. \par

The application of dielectric spectroscopy for food quality assessment has been widely investigated on many different crops, including wheat~\cite{TIM2,m1}, shelled peanuts~\cite{m2}, onions~\cite{m3}, apples~\cite{m4}, melons~\cite{m5}, and potatoes~\cite{p1,p2,p3,p4}. Previous studies have demonstrated that the dielectric properties of most crops are highly sensitive to their dry matter or moisture content ~\cite{TIM1,mo1,mo2,mo3,mo4,TIM3}. Several studies have specifically examined the dielectric properties of potatoes with different dry matter or moisture content for various microwave applications~\cite{p1,p2,p3,p4}. In~\cite{p1}, the dielectric properties of raw potatoes and potato chips with different moisture contents were measured using the slotted line technique at 300, 1000, and 3000 MHz at 25°C, aiming to identify suitable frequencies for microwave finish drying of potato chips. In~\cite{p2}, the dielectric properties of sorbed freeze-dried potatoes at 3 GHz and 25°C were investigated using standing wave measurements. It provided insights into the influence of moisture content and water activity on the dielectric behavior of semi-solid food. However, the investigated sorbed freeze-dried samples differ from fresh potatoes in their natural state. In~\cite{p3}, the change in dielectric properties between fresh and thawed potatoes was examined from 500 MHz to 20 GHz using an open-ended coaxial probe. The study demonstrated that dielectric spectroscopy could be a promising non-destructive method for detecting freezing and thawing damage. In~\cite{p4}, dielectric properties of raw potato tubers and simplified agar gel-based samples with controlled dry matter content were measured across the frequency range of 0.01 GHz to 3 GHz using an open-ended coaxial probe. The dry matter content of raw potato was calculated based on the SG. Both linear model and multivariate partial least squares regression model were developed to establish the relationship between dielectric properties and dry matter content. Although the models exhibited high accuracy in homogeneous agar gel-based samples, their performance on raw potatoes was limited, with the coefficient of determination (\(R^{2}\)) of only 0.57. The reduced performance is primarily due to the measurement setup not accounting for the spatial heterogeneity of starch distribution in potatoes. Further research is required to develop an accurate model that can reliably link the dielectric properties of potatoes with their SG. \par

\begin{figure}[t]
	\centering
	\centerline{\includegraphics[width=1.0\linewidth]{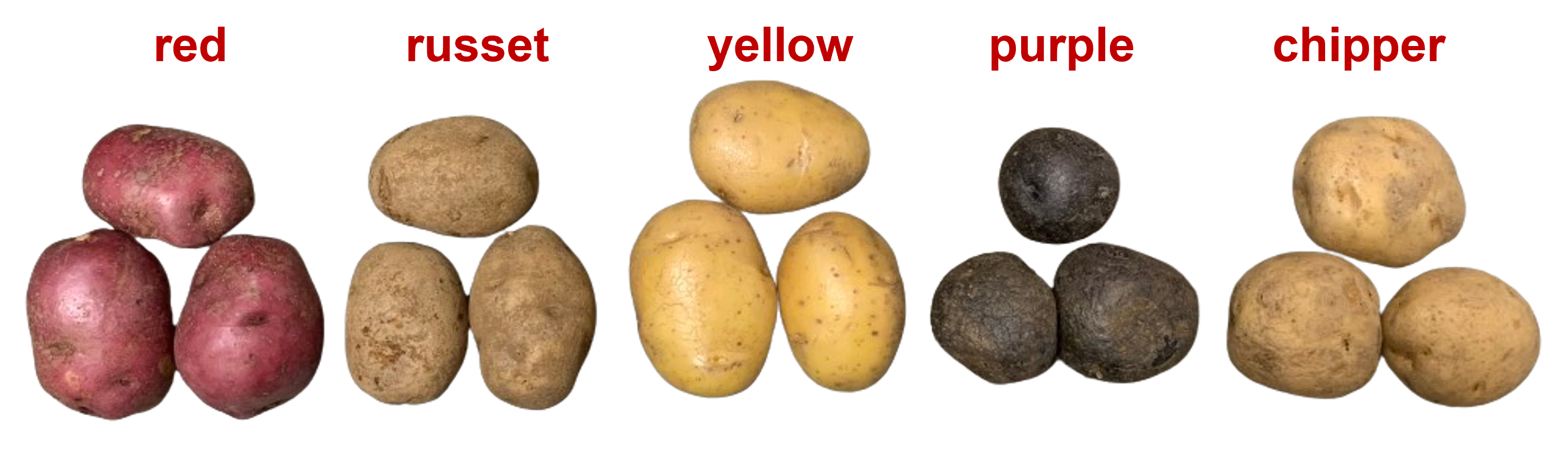}}
	\caption{The five types of potatoes for measurements. }
	\label{fig:harvestpotato}
\end{figure}

In this study, we address this gap by building a robust model to estimate the SG of raw potatoes based on dielectric properties at any frequencies from 0.3 GHz to 3.0 GHz. A comprehensive set of dielectric spectroscopy and SG measurements was conducted on 250 freshly harvested potatoes from five different types, which are shown in Fig.~\ref{fig:harvestpotato}. Out of the 250 sets of measured data, 200 sets were used to develop the SG estimation model, and the remaining 50 sets were used for model testing. To account for the inhomogeneous distribution of dry matter and moisture, the dielectric properties were measured at three different locations on each potato and averaged to obtain the representative values. Based on the measured data, a linear regression model between the dielectric constant and SG was established, with the regression coefficients modeled as fourth-order polynomial functions of frequency to account for the frequency dispersion. This model allows SG estimation using the dielectric constant measured at any frequency between 0.3 GHz and 3.0 GHz. The model’s accuracy was validated on 50 testing samples of five types and 50 yellow potatoes at different growing stages, achieving mean absolute error (MAE) \(<4.8\times10^{-3}\) and mean absolute percentage error (MAPE) \(< 0.45\%\). The high estimation accuracy demonstrates the model’s robustness for SG estimation. The model provides a foundation for the future development of microwave sensing technologies for agronomic trait assessment of potatoes.  \par

The remainder of this paper is organized as follows. Section II describes the methods used for SG and dielectric spectroscopy measurements, and presents the corresponding results. Section III presents the derivation and validation of the SG estimation model. Section IV discusses measurement uncertainty, repeatability, the influence of temperature variation on the dielectric constant of potatoes, frequency selection, and future applications of the model. Finally, the conclusion is drawn in Section V. \par

\begin{figure}[!t]
    \centering
    \begin{tabular*}{\linewidth}{@{\hspace{1em}}c@{\hspace{1em}}c@{}}
         \includegraphics[width=0.38\linewidth]{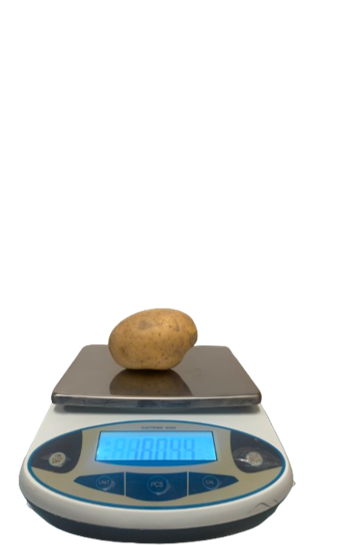} & 
         \includegraphics[width=0.48\linewidth]{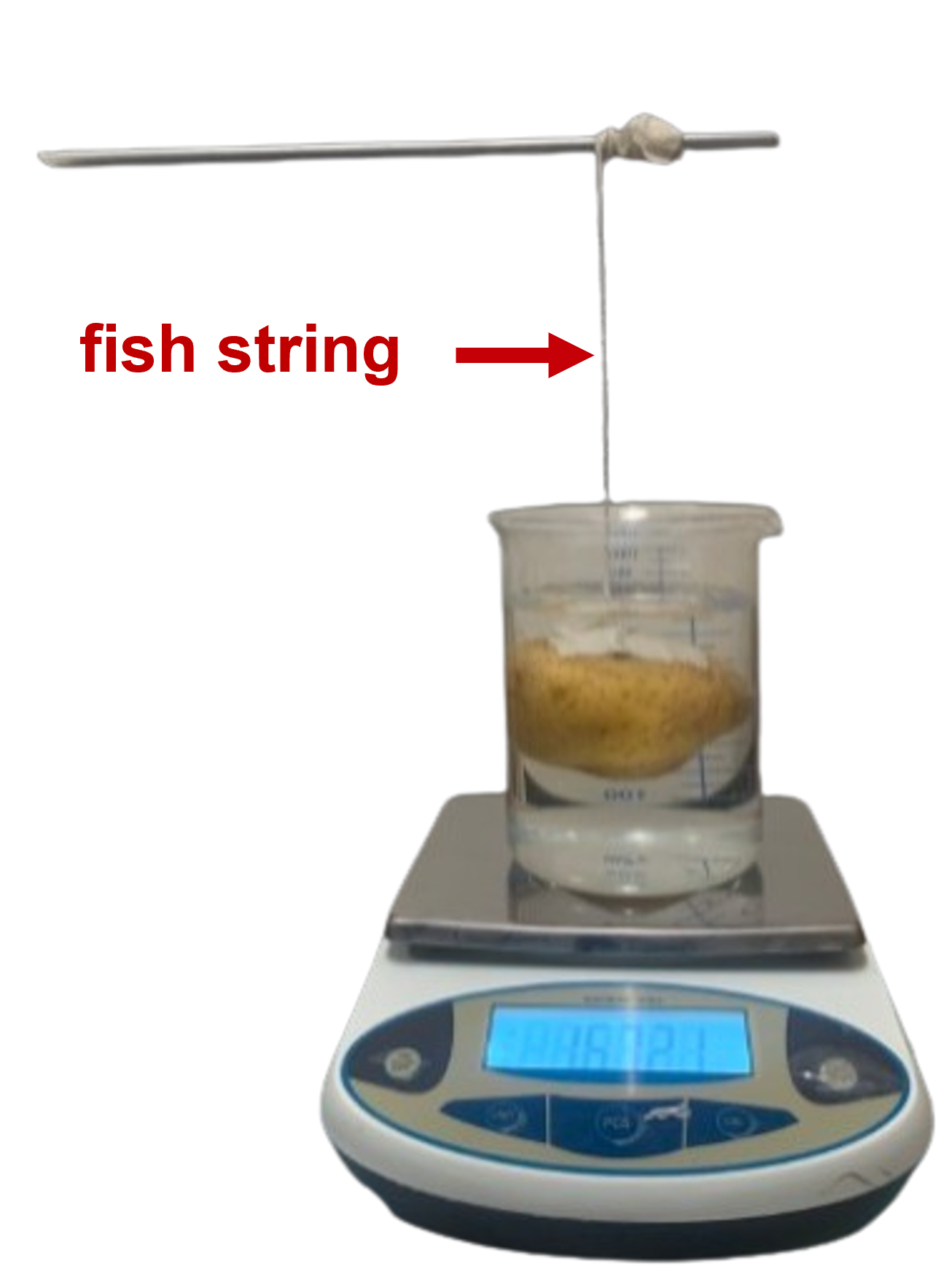} \\
         \footnotesize{(a)} & \hspace{4em}\footnotesize{(b)}
    \end{tabular*}
    \caption{Specific gravity (SG) measurement setup. (a) Measurement of \(W_{\mathrm{air}}\). (b) Measurement of \(F_{\mathrm{float}}\).}
    \label{fig:setup1}
\end{figure}

\section{Experimental Methods}\label{sec2}
The SG and dielectric properties of 250 potatoes were measured to build a comprehensive dataset for investigating the relationship between them. The 250 potatoes belong to five types associated with five different market sectors: red, russet, yellow, purple, and chipper, with 50 samples each. These types naturally have slightly different SG ranges, which allows the dataset to cover a broad SG spectrum. The collection, sampling and storage condition for the tested potatoes tubers are presented in Subsection A. The measurement methods of the SG and dielectric properties are presented in Subsections B and C, and the experimental results are presented in Subsection D.\par

\subsection{Sampling and Storage}

Freshly harvested russet, red, yellow, purple, and chipping potatoes were collected from the Hancock Agricultural Research Station, University of Wisconsin–Madison. Following harvest, the potatoes were stored under controlled conditions: red, yellow, and purple potatoes at 4.4 °C, russet potatoes at 8 °C, and chipping potatoes at 6 °C. Storage humidity was maintained at 95\%, and all potatoes were treated with a sprout inhibitor in accordance with industry protocols. The potatoes were then transported from the research station to the laboratory within approximately 2 hours and stored in a constant-temperature environment (7°C) with 95\% relative humidity.\par

Prior to measurement, potatoes were taken out of storage and left at room temperature for 2–3 hours until they reached room temperature (19-24°C). SG and dielectric measurements were then performed. All transported samples were measured within 7 days of collection to minimize quality changes.\par

Regarding sampling size, we followed the ISO 874:1980 standard~\cite{sample1}, which requires at least 3 kg of tubers to represent a tuber type. In our study, we randomly selected tubers of each type (red, russet, yellow, purple, and chipping potatoes)~\cite{sample2}. The accumulated total weights per potato type were: red potatoes, over 6.65 kg; russet potatoes, over 7.31 kg; yellow potatoes, over 7.10 kg; purple potatoes, over 4.61 kg; and chipping potatoes, over 6.06 kg. These values exceed the minimum ISO 874:1980 standard requirement, which ensures that the sampling was representative for each potato type.\par

\subsection{Specific Gravity (SG) Measurements}
We used a modified flotation method to measure the SG of potatoes. The traditional flotation method~\cite{r24} for SG measurement is:

\begin{equation}
\label{equ_1}
\mathrm{SG} = \frac{W_{\mathrm{air}}}{W_{\mathrm{air}} - W_{\mathrm{uww}}},
\end{equation}
where \(W_{\mathrm{air}}\) is the weight of a potato in air, and \(W_{\mathrm{uww}}\) is the weight of a potato when fully submerged in water.\par 

When the potato is completely submerged and at equilibrium, the underwater weight (\(W_{\mathrm{uww}}\)) is equal to the weight in air (\(W_{\mathrm{air}}\)) minus the buoyant force (\(F_{\mathrm{float}}\)) acting on it. Therefore, Eq.~\eqref{equ_1} can be rewritten as:\par
\begin{equation}
\label{equ_3}
\mathrm{SG} = \frac{W_{\mathrm{air}}}{F_{\mathrm{float}}}.
\end{equation}

In this work, we used the measurement setup shown in Figs.~\ref{fig:setup1}(a) and ~\ref{fig:setup1}(b) to measure \(W_{\mathrm{air}}\) and \(F_{\mathrm{float}}\) for each potato. A Bonvoisin Lab Scale with a precision of 0.01 g was used. The potato was first weighed in air to obtain \(W_{\mathrm{air}}\). Next, as shown in Fig.~\ref{fig:setup1}(b), a plastic beaker filled with distilled water was placed on the scale, and the scale was tared to cancel out the weight of the beaker and water. The potato was then fully submerged and stabilized using a thin fish string. The reading on the scale corresponds to \(F_{\mathrm{float}}/g\), where \(g\) is the gravitational acceleration. The SG was then calculated using Eq.~\eqref{equ_3} based on the measured values of \(W_{\mathrm{air}}\) and \(F_{\mathrm{float}}\). \par

\subsection{Dielectric Spectroscopy Measurements}

\begin{figure}[!t]
	\begin{center}
		\begin{tabular}{c@{ }}
			\includegraphics[width=0.8\linewidth]{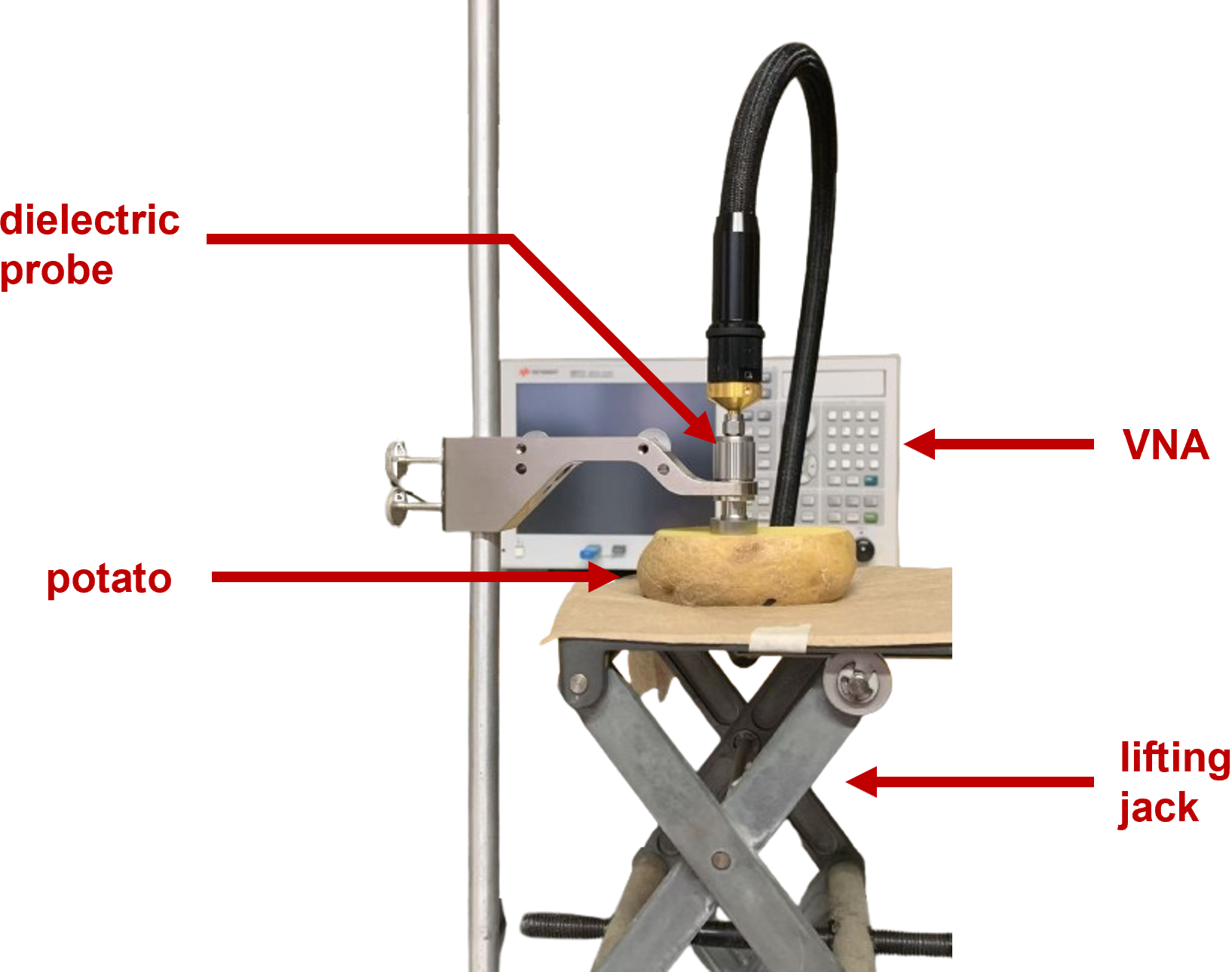}\\
			\footnotesize{(a)} \\
			\includegraphics[width=1.0\linewidth]{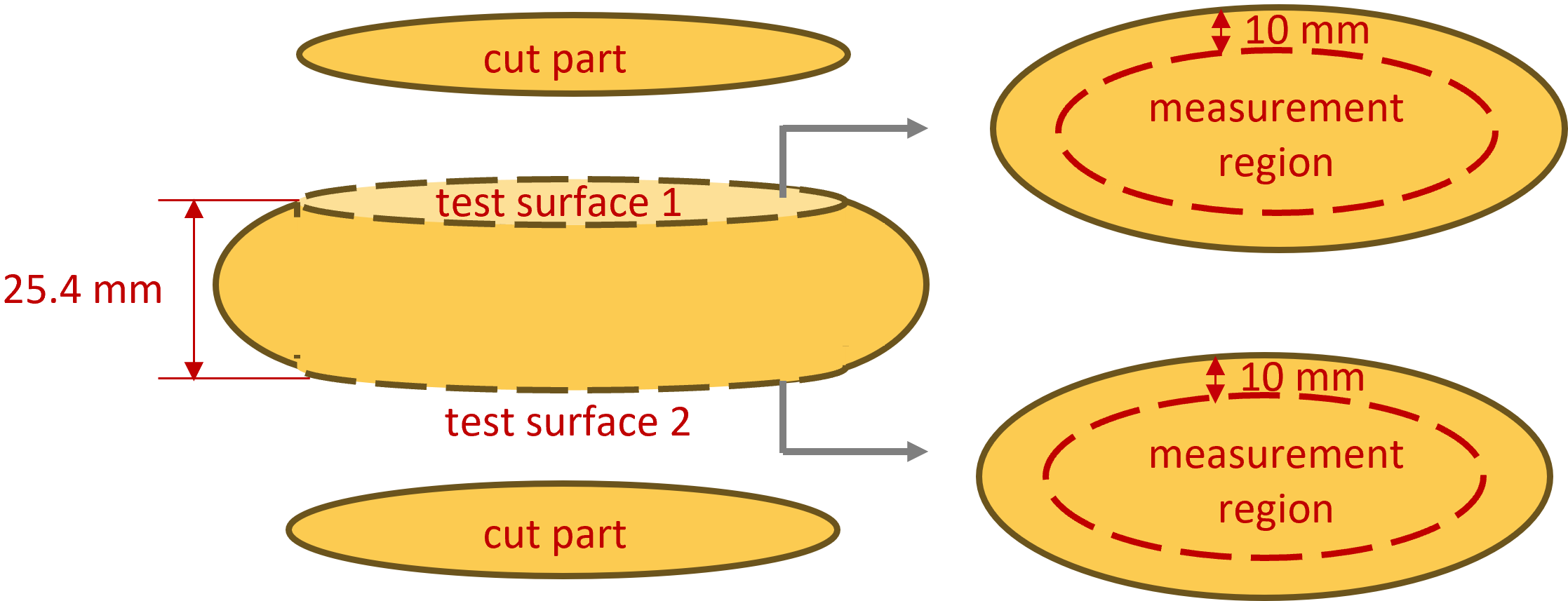}\\
			\footnotesize{(b)}  \\
		\end{tabular}
	\end{center}
	\caption{(a) Experimental setup for dielectric spectroscopy measurement of a potato sample. (b) Schematic view of the prepared tuber sample. The tuber sample is cut to create two parallel flat surfaces with their shortest diameter larger than 50 mm. The sample thickness is set to 25.4 mm. Measurements are taken on the two surfaces at locations at least 10 mm from the epidermis.}
	\label{fig:setup2}
\end{figure}

Dielectric spectroscopy measurements of potatoes were conducted over the frequency range from 0.3 to 3.0 GHz using a Keysight N1501A dielectric probe kit with a high-temperature open-ended coaxial probe. The frequency range was selected based on the common frequency spectrum used for underground crop detection and measurements~\cite{mics1,mics2,mics3,mics4,rGPR}. Theoretical support for this frequency range is discussed in detail in Section IV.D. The measurement setup is shown in Fig. \ref{fig:setup2}. The probe was mounted on a stand holder and connected to the Keysight E5071C vector network analyzer (VNA) via a phase-stable coaxial cable. Each potato sample is prepared based on the measurement requirement of the Keysight high temperature dielectric probe~\cite{rKSDoc}. Fig. 3(b) shows the schematic view of the prepared sample. We avoid the tissues such as the bud eyes and bud roots, and cut the tuber to create two smooth, flat, parallel surfaces for the measurement. Each surface has its shortest diameter larger than 50 mm, and the sample thickness is set to 25.4 mm. Measurements are taken on both surfaces at locations at least 10 mm from the epidermis. To ensure measurement accuracy, the surfaces must be clean and free of any coverings. The sample was placed on a manually adjustable lifting jack so that the sample position could be precisely adjusted to ensure full contact between the potato surface and the probe. This setup minimizes air gaps and improves measurement accuracy.  All measurements were performed at room temperature, with the internal temperature of the potato samples ranging from 19.0°C to 24.0°C, as measured using a temperature probe. The VNA output power was set to 0 dBm, and the intermediate frequency bandwidth (IFBW) was set to 1 kHz. A total of 283 frequency points from 0.3 GHz to 3.0 GHz were collected for each dielectric spectroscopy measurement.  \par

Given the inhomogeneous distribution of starch and moisture in potatoes, the complex relative permittivity ($\epsilon_r=\epsilon_r'-j\epsilon_r''$) was measured at three different locations on each sample to account for the variability. The three sets of measured data were then averaged to obtain the final relative permittivity for each potato.
\par

\begin{figure}[!t]
	\begin{center}
		\begin{tabular}{c@{ }}
			\includegraphics[width=0.7\linewidth]{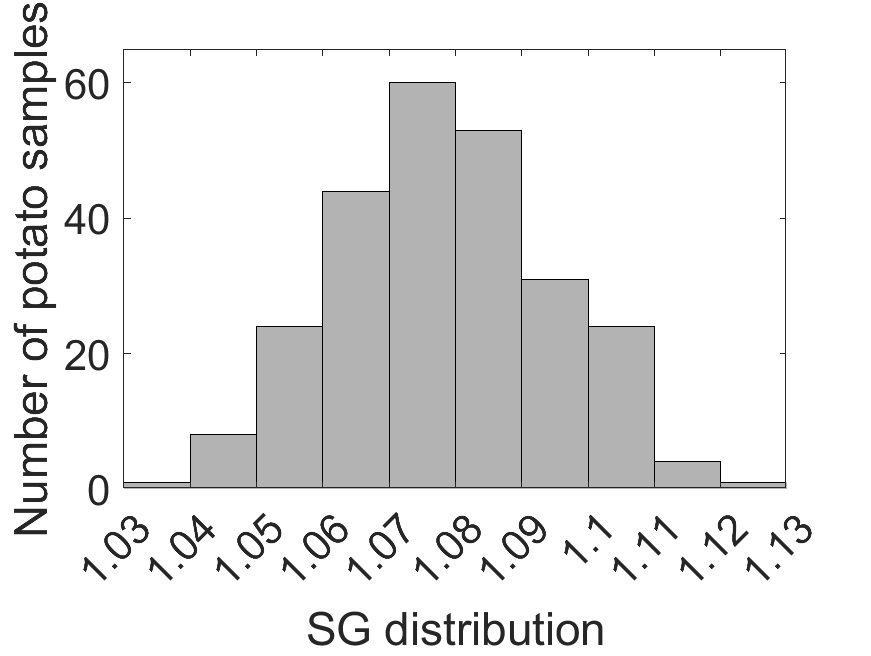}\\
			\footnotesize{(a)} \\
			\includegraphics[width=0.7\linewidth]{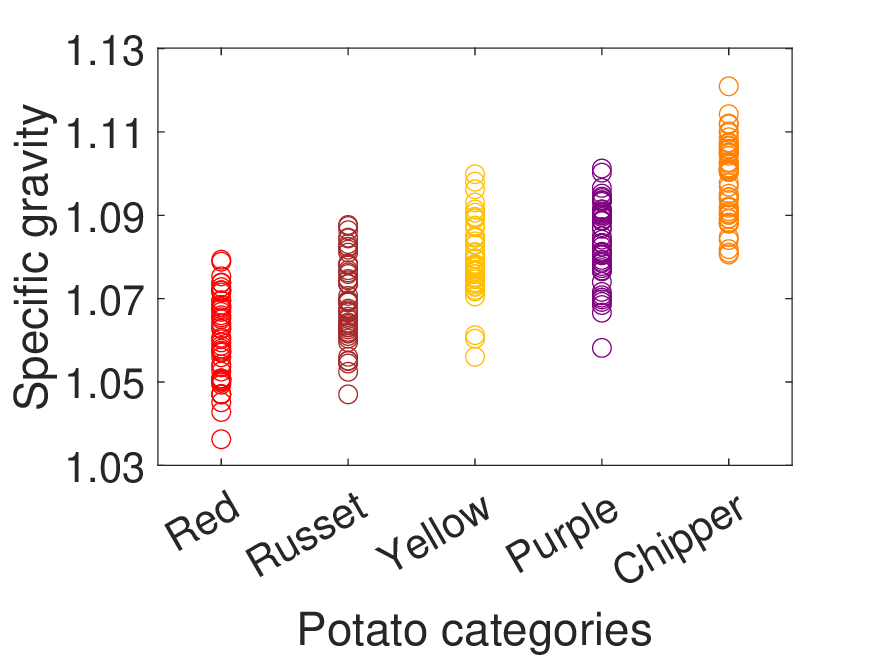}\\
			\footnotesize{(b)}  \\
		\end{tabular}
	\end{center}
	\caption{The measurement results of SG of the 250 potatoes (five types, 50 potatoes each). (a) Distribution of measured SG of the 250 potatoes. (b) Distribution of SG of potatoes from five different types.}
	\label{fig:SG}
\end{figure}

\subsection{Measurement Results and Analysis}
Fig. \ref{fig:SG}(a) shows the SG distribution of the 250 potato samples. The SG values ranges from 1.036 to 1.121. This range effectively covers the SG typically observed throughout the potato growth cycle, as well as across different market sectors in the potato industry (fresh market, processing, seed, and specialty sectors)~\cite{SGrange1,SGrange2}. Fig. \ref{fig:SG}(b) shows the SG distributions for the five types. Each type exhibits a slightly different SG range. Red potatoes fall within the lower SG range (1.036–1.079), while chipping potatoes span the higher SG range (1.081–1.121). The russet, yellow, and purple potatoes fall in between: russet (1.047–1.088), yellow (1.056–1.100), and purple (1.058–1.101). The continuous SG distribution within each type supports the development and evaluation of SG estimation models, as discussed in Section III.

\begin{figure}[t]
	\centering
	\centerline{\includegraphics[width=1\linewidth]{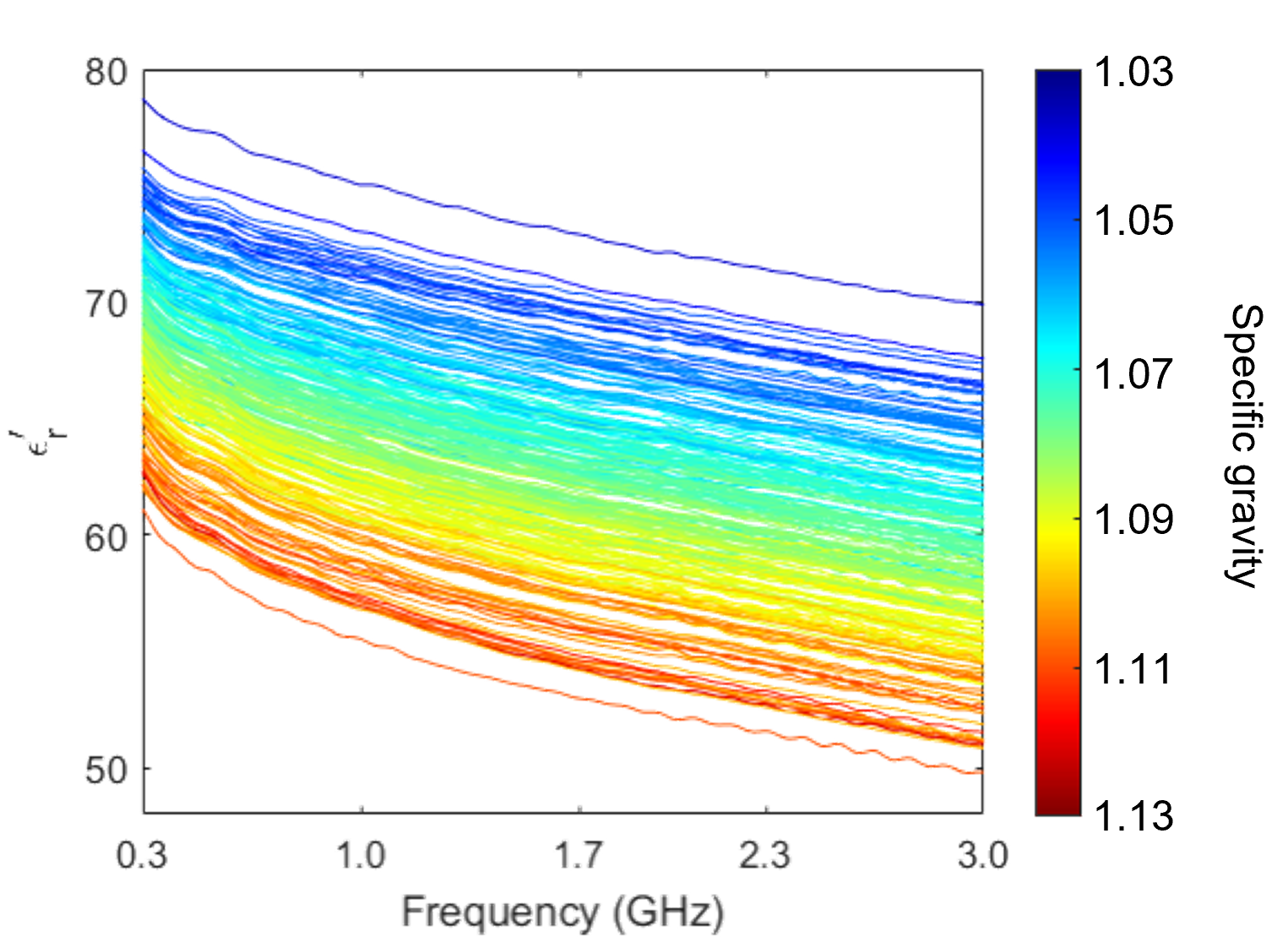}}
	\caption{The $\epsilon_r'$ values of the 250 potatoes from 0.3 GHz to 3.0 GHz. The transition of line color from blue to red represents increasing SG. }
	\label{fig:realpart}
\end{figure}

Fig. \ref{fig:realpart} shows the real part of the measured  relative  permittivity ($\epsilon_r'$), also known as dielectric constant, of the 250 potatoes from 0.3 GHz to 3.0 GHz. The transition of line color from blue to red represents increasing SG values. For each potato sample, $\epsilon_r'$ decreases with increasing frequency, reflecting the dispersive properties of potato tissue. At each frequency, $\epsilon_r'$ decreases as SG increases. This inverse relationship is attributed to the dry matter-to-water ratio in potatoes: dry matter has a lower $\epsilon_r'$ than water~\cite{rstarch}, therefore, higher SG, which corresponds to a higher dry matter-to-water ratio, resulting in a lower $\epsilon_r'$. \par

\begin{figure}[!t]
	\centering
	\centerline{\includegraphics[width=1\linewidth]{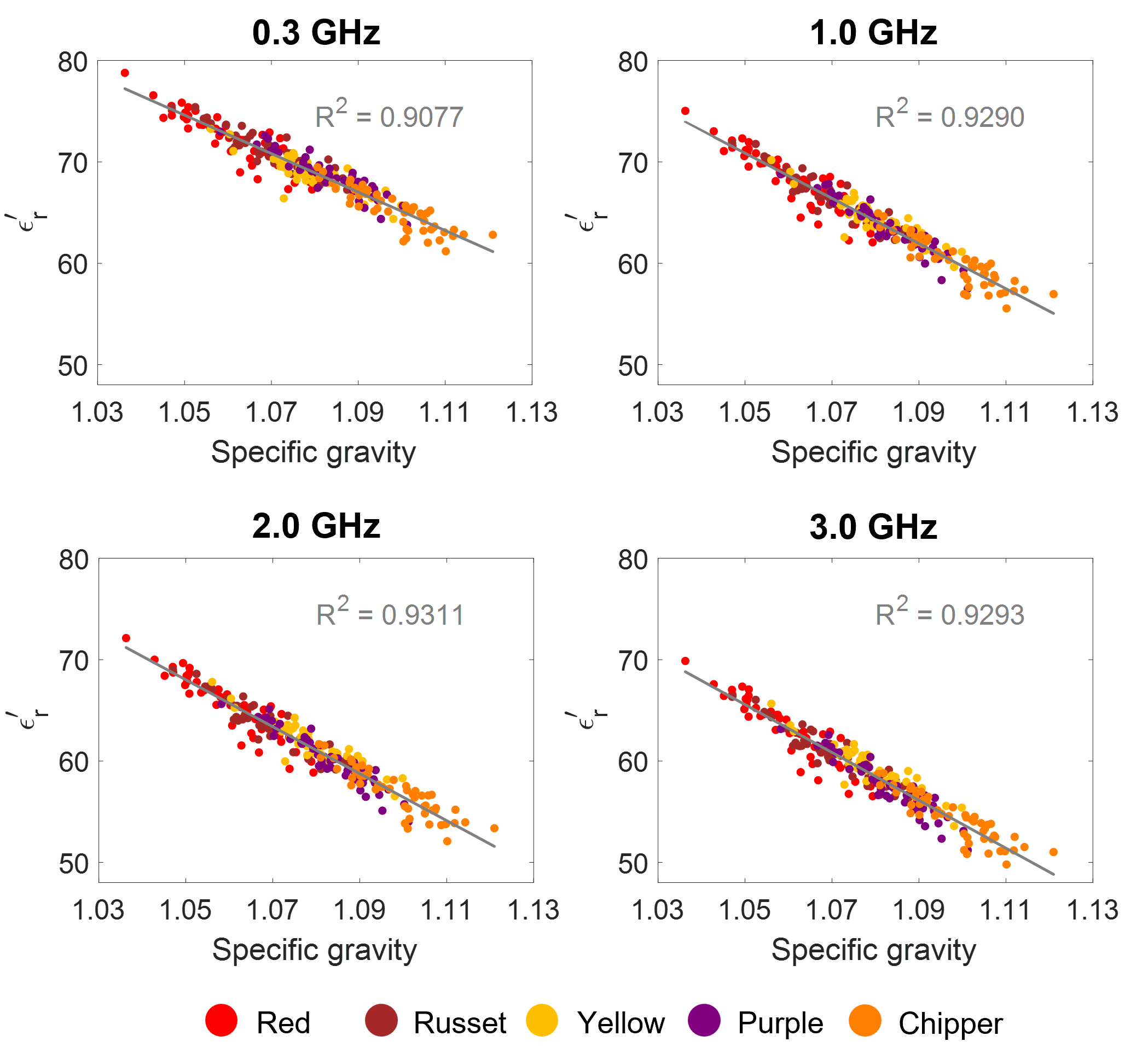}}
	\caption{The variation of $\epsilon_r'$ with SG at 0.3 GHz, 1.0 GHz, 2.0 GHz, and 3.0 GHz. Each data point is colored according to its type, where red, brown, yellow, purple, and orange correspond to red, russet, yellow, purple, and chipping potatoes, respectively.}
	\label{fig:real8freq}
\end{figure}

To better evaluate the relationship between $\epsilon_r'$ and SG, the $\epsilon_r'$ values at four frequencies (0.3 GHz, 1.0 GHz, 2.0 GHz, and 3.0 GHz) are compared with their corresponding SG values, as shown in Fig. \ref{fig:real8freq}, where data points in different colors represent different potato types. Although the dielectric constant varies in different ranges at different frequencies, a strong linear relationship is observed between SG and $\epsilon_r'$ at each frequency with \(R^{2}\) higher than 0.90. This consistency suggests that a linear model can be established between SG and $\epsilon_r'$ at any selected frequency. Moreover, the linear relationship is consistent across all five types of potatoes, indicating that the SG - $\epsilon_r'$ relationship is primarily attributed to the dry matter-to-water ratio, rather than being specific to a particular potato type.
\par

\begin{figure}[!t]
	\centering
	\centerline{\includegraphics[width=1\linewidth]{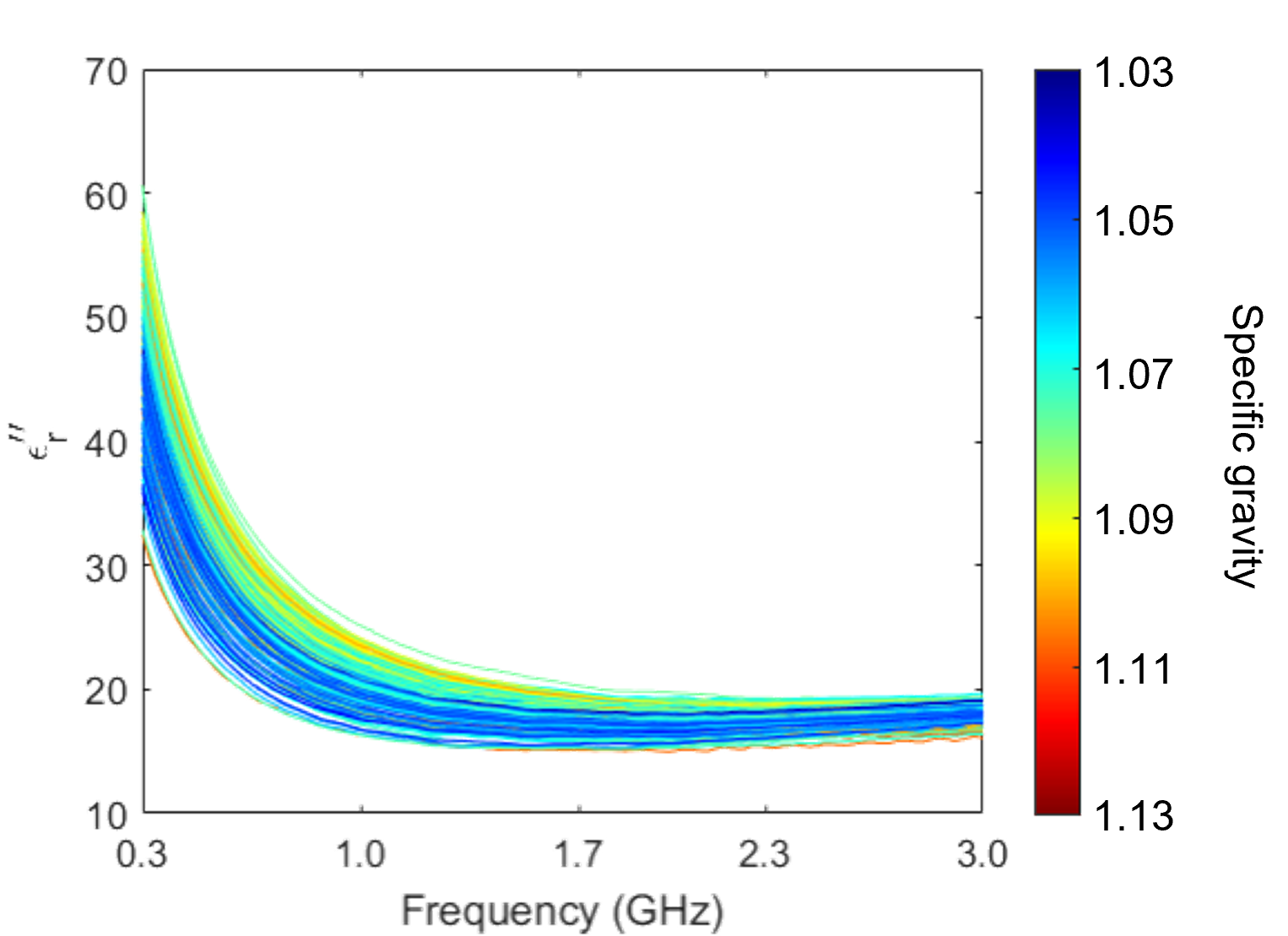}}
	\caption{The $\epsilon_r''$ values from 0.3 GHz to 3.0 GHz for 250 potatoes. The transition of line color from blue to red represents increasing SG.}
	\label{fig:imagpart}
\end{figure}

\begin{figure}[!t]
	\centering
	\centerline{\includegraphics[width=1\linewidth]{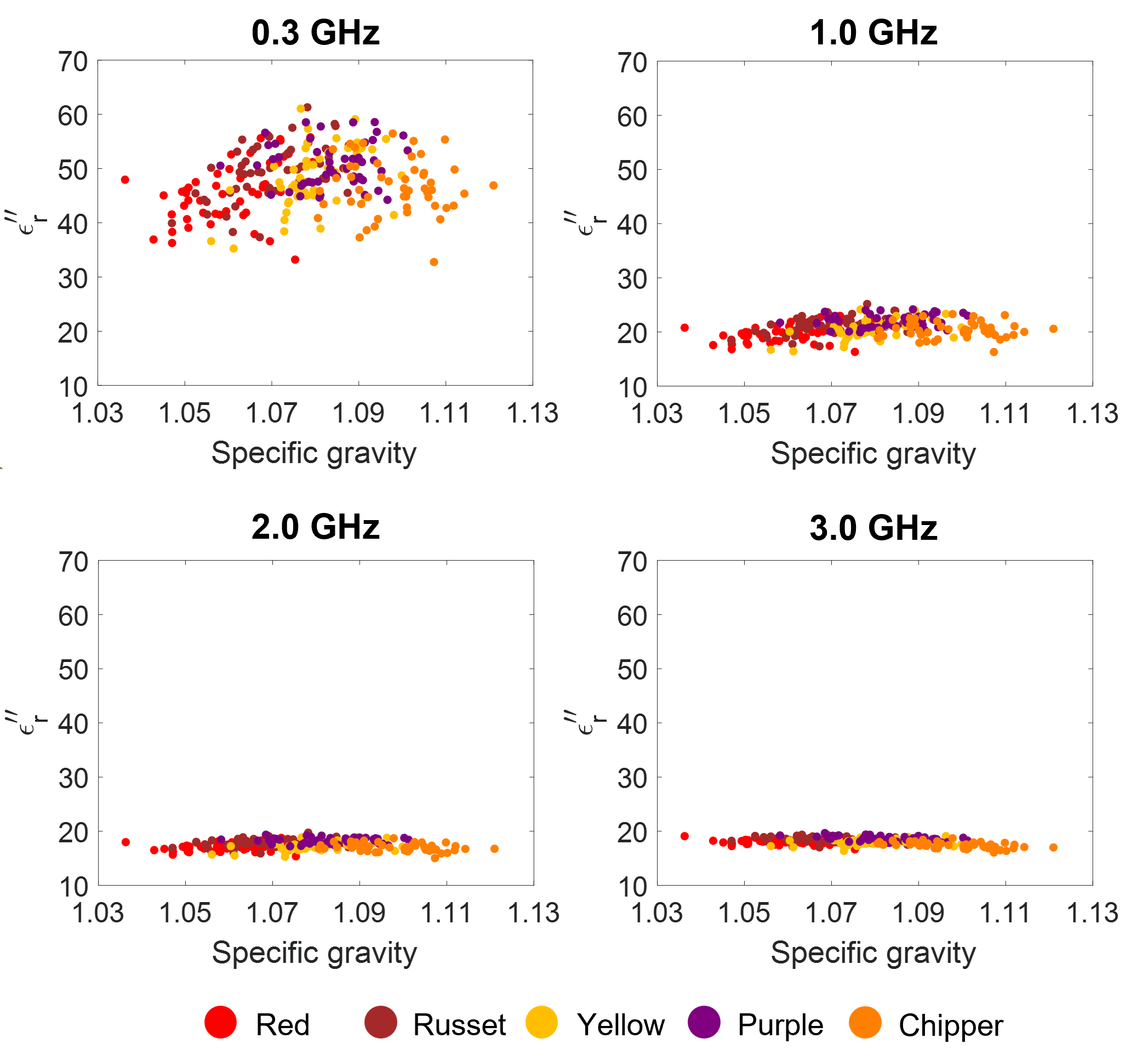}}
	\caption{The variation of $\epsilon_r''$ with SG at 0.3 GHz, 1.0 GHz, 2.0 GHz, and 3.0 GHz. Each data point is colored according to the potato type.}
	\label{fig:imag8freq}
\end{figure}

The imaginary part of the measured relative permittivity ($\epsilon_r''$) of all 250 potato samples from 0.3 GHz to 3.0 GHz is shown in Fig. \ref{fig:imagpart}. The transition of line color from blue to red represents increasing SG values. For each potato sample, $\epsilon_r''$ decreases nonlinearly with increasing frequency, with a steeper decrease at lower frequencies and a nearly constant trend beyond 1.5 GHz. At each frequency point, $\epsilon_r''$ is the lowest for low-SG potatoes (blue lines, SG = 1.03–1.07), increases for medium-SG potatoes (green lines, SG = 1.07–1.09), and then decreases for high-SG potatoes (red lines, SG = 1.09–1.13). This non-monotonic trend is more clearly illustrated at selected frequencies shown in Fig. \ref{fig:imag8freq}. Due to this non-monotonic relationship between $\epsilon_r''$ and SG, $\epsilon_r''$ is not suitable for developing an SG estimation model. The non-monotonic relationship between $\epsilon_r''$ and SG also negatively affects the relationship between $|\epsilon_r|$ and SG, making $|\epsilon_r|$ a less robust parameter for SG estimation. \par

In summary, the analysis of the dielectric properties of the 250 potato samples with different SG values shows that the dielectric constant $\epsilon_r'$ exhibits a consistent linear relationship with SG, whereas  $\epsilon_r''$ shows a non-monotonic relationship. Therefore, $\epsilon_r'$ is identified as a suitable electromagnetic parameter for SG estimation. The development and validation of the SG estimation model based on $\epsilon_r'$ are presented in Section III.
\par

\section{SG Estimation Model}\label{sec3}

We developed a linear regression model to estimate SG based on $\epsilon_r'$ and evaluated its accuracy using the measured data of the 250 potato samples. Each data set corresponding to a potato sample includes the SG value and $\epsilon_r'$ values at 283 frequency points from 0.3 GHz to 3.0 GHz. Among the 250 sets of data from five potato types, 200 (40 per potato type) were randomly selected for model development and are referred to as the training data. The remaining 50 (10 per type) were used to validate the model's accuracy and are referred to as the testing data. \par

\begin{figure}[!t]
	\begin{center}
		\begin{tabular}{c@{ }}
			\includegraphics[width=0.8\linewidth]{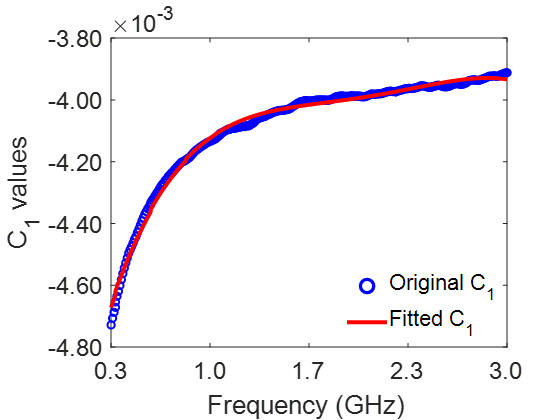}\\
			\footnotesize{(a)} \\
			\includegraphics[width=0.8\linewidth]{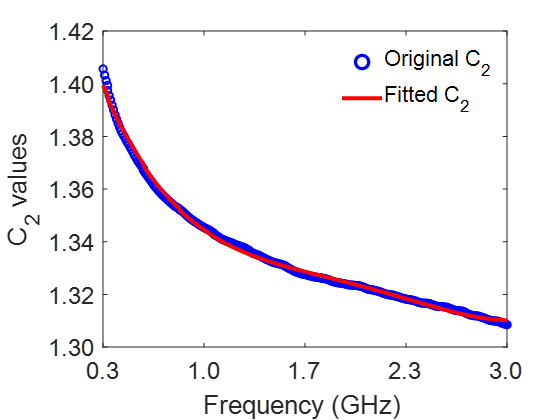}\\
			\footnotesize{(b)}  \\
		\end{tabular}
	\end{center}
	\caption{Original and fitted (a) \(C_1\) and (b) \(C_2\) coefficients from 0.3 GHz to 3.0 GHz. }
	\label{fig:Cprefull}
\end{figure}

\subsection{Specific Gravity (SG) Estimation Model}

At a specific frequency $f$, a linear regression model is used to establish the relationship between the dielectric constant at that frequency $\epsilon_r'(f)$ and the SG:

\begin{equation}
\label{equ_8}
{\rm SG}=C_{1}×\epsilon_r'(f)+C_{2}.
\end{equation}
The coefficients \(C_{1}\) and \(C_{2}\) are obtained by fitting the model to the 200 training data sets and minimizing the least-squares error using MATLAB's \texttt{polyfit} function.  Due to the dispersive nature of potato tissue, at each frequency point, the model has a unique pair of \{\(C_{1}\), \(C_{2}\)\}. The fitted \(C_{1}\) and \(C_{2}\) values at the 283 measured frequency points from 0.3 GHz to 3.0 GHz are shown in Fig. \ref{fig:Cprefull}. We refer to these values as the original \(C_{1}\) and \(C_{2}\) values. Both coefficients show nonlinear variation with frequency. 

To characterize the frequency dependence of \(C_1\) and \(C_2\), a fourth-order polynomial model was used: 

\begin{equation}
\label{equ_9}
{C(f)}=a_{0} + a_{1}f + a_{2}f^{2} + a_{3}f^{3} + a_{4}f^{4}.
\end{equation}
Here, \(C\) represents either \(C_1\) or \(C_2\), and $f$ is the frequency measured in GHz. The coefficients \{\(a_{0}\), \(a_{1}\), ..., \(a_{4}\)\} were obtained using a MATLAB fitting function that minimizes the least-squares error. The resulting coefficients for both \(C_1\) and \(C_2\) are summarized in Table I. The fitted curves, shown as solid lines in Fig.~\ref{fig:Cprefull}, demonstrate excellent agreement with the original values.

Eqs.~\eqref{equ_8} and ~\eqref{equ_9} are combined to form the SG estimation model. Given the dielectric constant at any specific frequency between 0.3 GHz and 3.0 GHz, Eq.~\eqref{equ_9} is used to calculate the \(C_1\) and \(C_2\) for that frequency, and then Eq.~\eqref{equ_8} uses these two coefficients and the dielectric constant to calculate the potato's SG value.\par

\begin{table}[tb]
	\caption{Coefficient for the Calculation of $C_{1}$ and $C_{2}$  }
	\centering
	\begin{tabular}{c@{\hskip 50pt} c@{\hskip 50pt} c}
		\hline
		\hline
		   & \textbf{$C_{1}$} & 
                \textbf{$C_{2}$}  \\
		\hline
\textbf{$a_{0}$} & $-5.222\times10^{-3}$ &  1.451 \\
\textbf{$a_{1}$} & $2.320\times10^{-3}$ &  -0.215 \\
\textbf{$a_{2}$} & $-1.707\times10^{-3}$ &  0.151 \\
\textbf{$a_{3}$} & $0.564\times10^{-3}$ &  -0.050 \\
\textbf{$a_{4}$} & $-0.068\times10^{-3}$ &  0.006 \\
		\hline
		\hline
	\end{tabular}
	\label{table:coef}
\end{table}

\subsection{Model Validation}

\begin{figure}[tb]
	\centering
	\centerline{\includegraphics[width=1\linewidth]{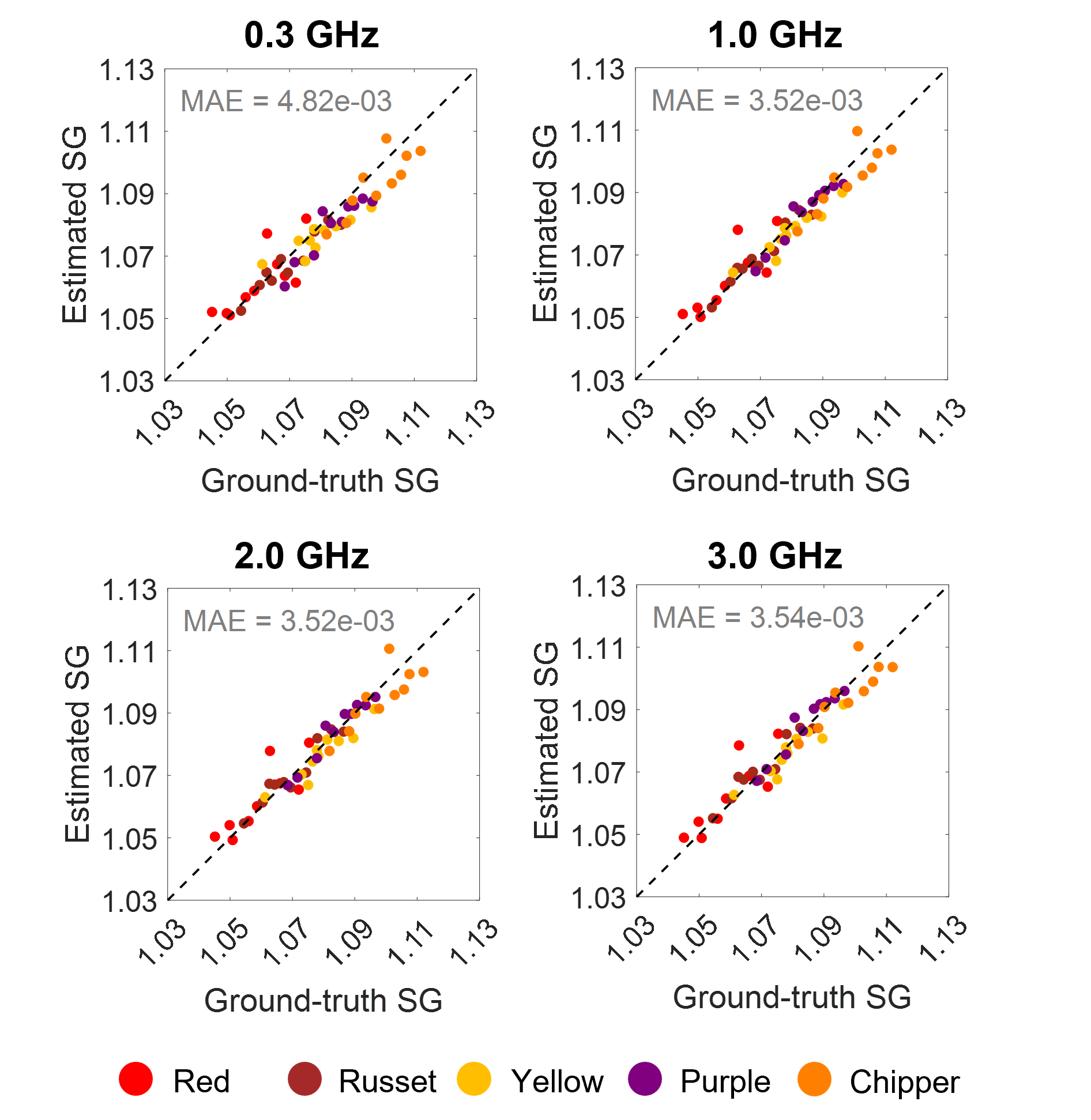}}
	\caption{The comparison of model-estimated SG with ground-truth SG for 50 testing potato samples at 0.3 GHz, 1.0 GHz, 2.0 GHz, and 3.0 GHz. Data points for different potato types are shown in different colors.}
	\label{fig:SGPredict}
\end{figure}

\begin{figure}[!t]
    \centering
    \begin{tabular*}{\linewidth}{@{\extracolsep{\fill}}cc}
         \includegraphics[width=0.48\linewidth]{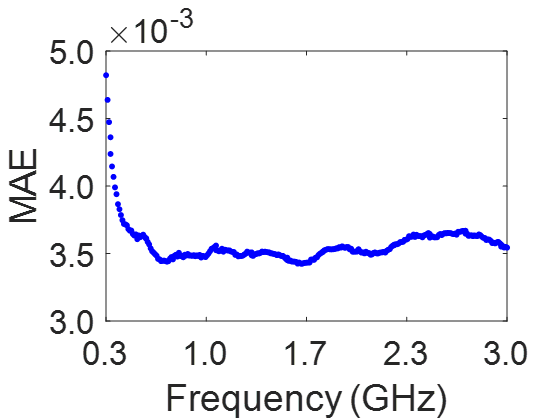} & 
         \includegraphics[width=0.48\linewidth]{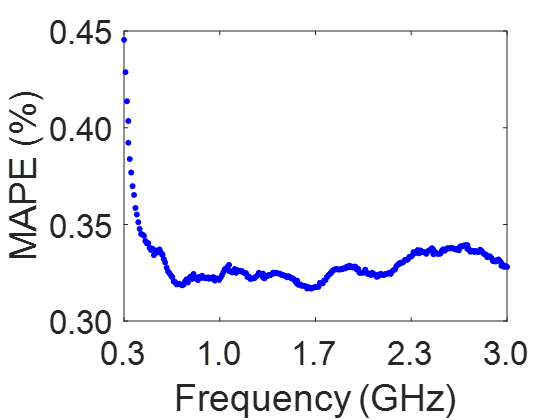} \\
         \footnotesize{(a)} & \footnotesize{(b)}
    \end{tabular*}
    \caption{(a) MAE, and (b) MAPE  between the model-estimated and ground-truth SG for the 50 testing potato samples from 0.3 GHz to 3.0 GHz.}
    \label{fig:GMAE}
\end{figure}

We evaluated the model's accuracy using the 50 testing data sets that were not included in model derivation. At each frequency, we used Eq.~\eqref{equ_9} to calculate the \(C_1\) and \(C_2\), and then applied these coefficients together with the measured dielectric constant of the testing samples to Eq.~\eqref{equ_8} to estimate the SG of the samples. The model-estimated SG values were compared with the measured (ground-truth) SG using mean absolute error (MAE) and mean absolute percentage error (MAPE). Fig. \ref{fig:SGPredict} shows the estimated SG and ground-truth SG for the 50 testing potato samples at four frequencies, 0.3 GHz, 1.0 GHz, 2.0 GHz, and 3.0 GHz. The dots of different colors represent different potato types. The estimated SG values align closely with the ground-truth SG values across all types. Figs. \ref{fig:GMAE}(a) and \ref{fig:GMAE}(b) show the MAE and MAPE between the  model-estimated and ground-truth SG across the frequency range. The model accurately estimates the SG of the 50 testing potatoes with MAE less than \(4.8\times10^{-3}\) and MAPE less than 0.45\% at all frequencies. The model shows slightly better accuracy at higher frequencies above 0.5 GHz. The averaged MAE and MAPE over all the 283 frequency points are \(3.57\times10^{-3}\) and 0.33\%, respectively, as summarized in Table~\ref{table:vmae}.  These low error values confirm the model's effectiveness in SG estimation. \par

\begin{figure}[tb]
	\centering
	\centerline{\includegraphics[width=1\linewidth]{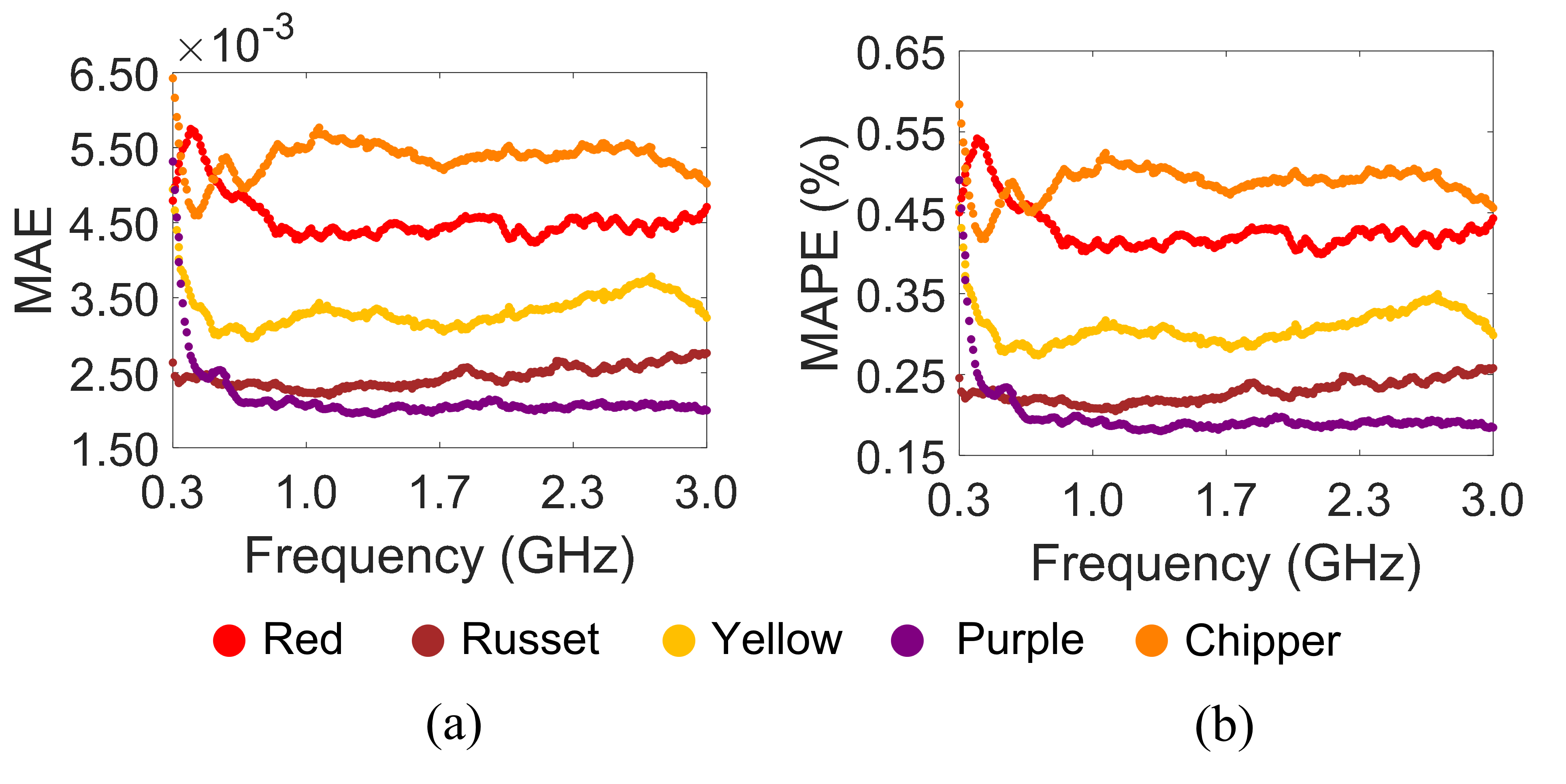}}
	\caption{(a) MAE, and (b) MAPE between the model-estimated and ground-truth SG for potatoes of five different types from 0.3 GHz to 3.0 GHz. Different types are represented by different colors.}
	\label{fig:VAMAE}
\end{figure}

To assess the model's performance across different potato types, the MAE and MAPE between the model-estimated and ground-truth SG were calculated separately for each potato type. The results are shown in Fig. \ref{fig:VAMAE}. The averaged MAE and MAPE over all the frequency points for each type are summarized in Table~\ref{table:vmae}. The model shows higher accuracy for russet, yellow, and purple potatoes with averaged MAE less than \(3.34\times10^{-3}\) and averaged MAPE less than 0.31\%, compared to red and chipping potatoes with averaged MAE less than \(5.37\times10^{-3}\) and averaged MAPE less than 0.49\%.  The slightly higher errors for red and chipping potatoes are attributed to their SG values being near the boundaries of the SG range used for model development, as shown in Fig.~\ref{fig:SG}. Nevertheless, the error metrics remain small, demonstrating the model’s generalizability for estimating SG of potatoes.

\begin{table}[t]
	\caption{Accuracy of the Model-Estimated SG for Different Potato Types}
	\centering
	\begin{tabular}{c@{\hskip 50pt} c@{\hskip 50pt} c}
		\hline
		\hline
		\textbf{Types}   & \textbf{$\overline{MAE}$} & \textbf{$\overline{MAPE}(\%)$}\\
		\hline
\textbf{Total} & $3.57\times10^{-3}$ & $0.33$ \\
\textbf{Red} & $4.55\times10^{-3}$ & $0.43$ \\
\textbf{Russet} & $2.45\times10^{-3}$ & $0.23$ \\
\textbf{Yellow} & $3.34\times10^{-3}$ & $0.31$ \\
\textbf{Purple} & $2.15\times10^{-3}$ & $0.20$ \\
\textbf{Chipper} & $5.37\times10^{-3}$ & $0.49$ \\
		\hline
		\hline
	\end{tabular}
	\label{table:vmae}
\end{table}

\subsection{Model Validation on Potato Tubers at Different Growing Stages}
\begin{figure}[!t]
    \centering
    \begin{tabular*}{\linewidth}{@{\extracolsep{\fill}}cc}
         \includegraphics[width=0.48\linewidth]{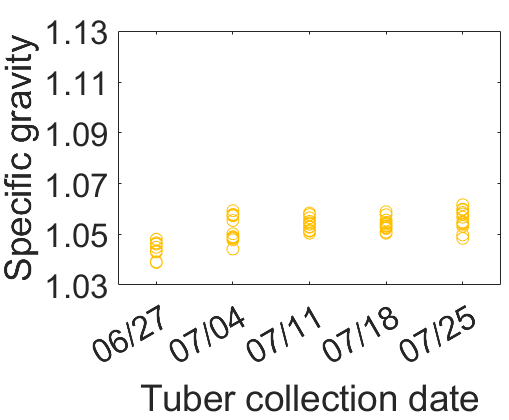} & 
         \includegraphics[width=0.38\linewidth]{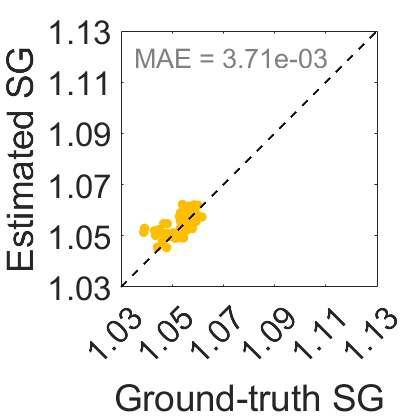} \\
         \footnotesize{(a)} & \footnotesize{(b)}
    \end{tabular*}
    \caption{(a) Measured SG of yellow potato tubers at different growing stages over five continuous weeks (06/27, early growth stage; 07/25, before harvest). (b) The comparison of model-estimated SG with ground-truth SG for 50 potato samples at different growing stages.}
    \label{fig:GS1}
\end{figure}

In the summer of 2025, we collected tubers (yellow potato variety, Colomba) at different stages of development over five continuous weeks at a commercial potato farm (Coloma Farms, Inc., Wisconsin, U.S.). Each week, 10 potatoes tubers were harvested and measured for their dielectric constant and SG. The variation of SG from 06/27 (early growing stage) to 07/25 (before harvested stage) are shown in Fig. \ref{fig:GS1}(a). As expected, SG increases slightly as the tubers mature. Based on the measured dielectric constant at 1.0 GHz, the SG of the 50 tuber samples is estimated using our model, and the results are shown in Fig. \ref{fig:GS1}(b). The estimated SG values align closely with the ground-truth SG values, with MAE of \(3.71\times10^{-3}\) and MAPE of 0.35\%. The estimation accuracy is comparable to that for yellow potatoes at maturity, which demonstrates the effectiveness of our model in characterizing SG at different stages of tuber development. \par

\section{Discussion}\label{sec4}
In this section, we present the standard deviation of the dielectric constant measurements, discuss measurement repeatability, examine the influence of temperature variation on the dielectric constant of potatoes, provide theoretical support for the selection of the frequency range used to build the SG estimation model, and discuss future applications of the model.

\subsection{Standard Deviation of the Dielectric Constant Measurements}
\begin{figure*}[tb]
	\centering
	\centerline{\includegraphics[width = 1\linewidth]{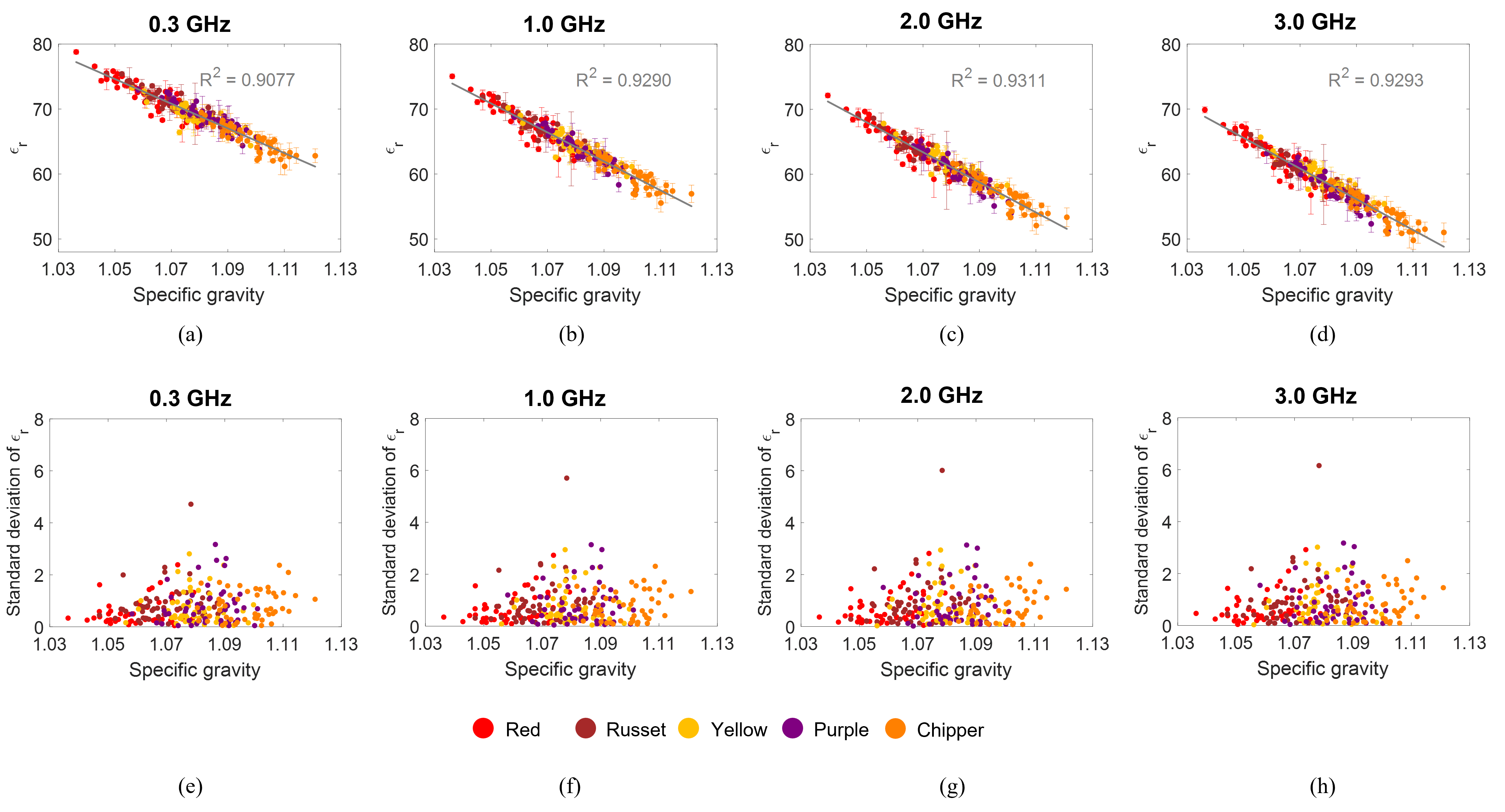}}
	\caption{(a) - (d) The variation of $\epsilon_r'$ with SG of potato samples at 0.3 GHz, 1.0 GHz, 2.0 GHz, and 3.0 GHz. Error bars represent the standard deviation of measurements at three positions on each sample. (e) - (h) The standard deviation of dielectric constant with different SG at 0.3 GHz, 1.0 GHz, 2.0 GHz, and 3.0 GHz. }
	\label{fig:SD}
\end{figure*}

The measured dielectric constant of each potato sample at four frequencies (0.3 GHz, 1.0 GHz, 2.0 GHz, and 3.0 GHz) is shown in Figs. \ref{fig:SD}(a) to \ref{fig:SD}(d). The standard deviation of measurements at three positions of each sample is shown as the error bar. The standard deviation of dielectric constant with different SG is shown in Figs. \ref{fig:SD}(e) to \ref{fig:SD}(h). In general, the standard deviation is larger for potatoes with SG in the mid-range (1.06–1.09) and smaller for those at both lower and higher SG values. This confirms the presence of internal heterogeneity in potatoes and suggests that tubers with low or high SG may have a more homogeneous internal structure compared to those with mid-range SG. Therefore, a multiple-position averaging approach is preferred when determining the representative dielectric constant of potatoes.\par

\subsection{Repeatability Analysis}
\begin{figure}[tb]
	\centering
	\centerline{\includegraphics[width=1\linewidth]{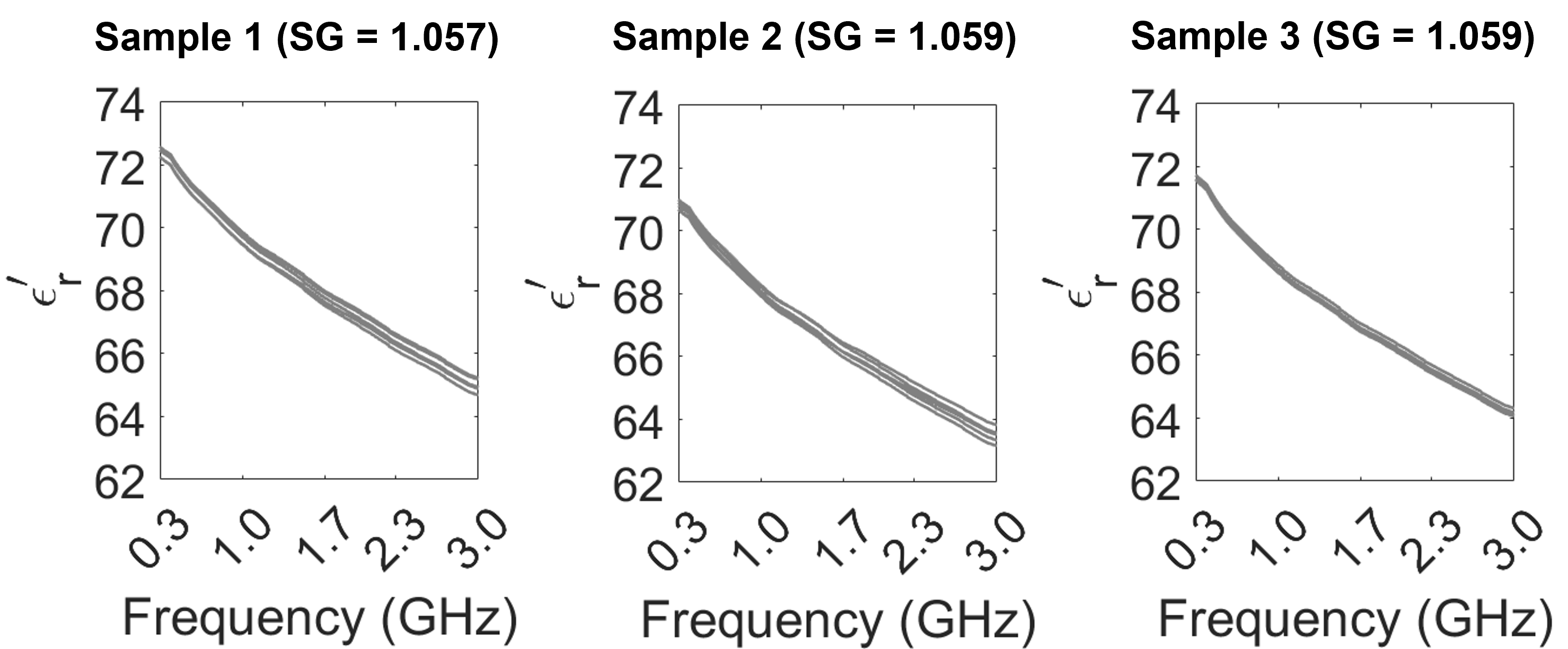}}
	\caption{Variation of the dielectric constant with five repeated measurements for three potato samples over the frequency range of 0.3–3.0 GHz. The line color represents SG.}
	\label{fig:Repeat}
\end{figure}

We performed repeatability measurements on three fresh potatoes. For each sample, following the same process described in the manuscript, we selected one spot and performed five repeated measurements by re-seating the probe at the same position. The results are shown in Fig.~\ref{fig:Repeat}. The maximum differences in dielectric constant across the frequency range are 0.6, 0.7, 0.3, demonstrating a high level of repeatability. Applying these dielectric constant variations due to repeated measurements to our SG estimation model leads to a MAE of less than \(2.76\times10^{-3}\), indicating that repeated measurements have a minimal impact on SG estimation accuracy.\par

\subsection{Influence of Temperature Variation on the Dielectric Constant of Potato Tubers}
\begin{figure}[tb]
	\centering
	\centerline{\includegraphics[width=1\linewidth]{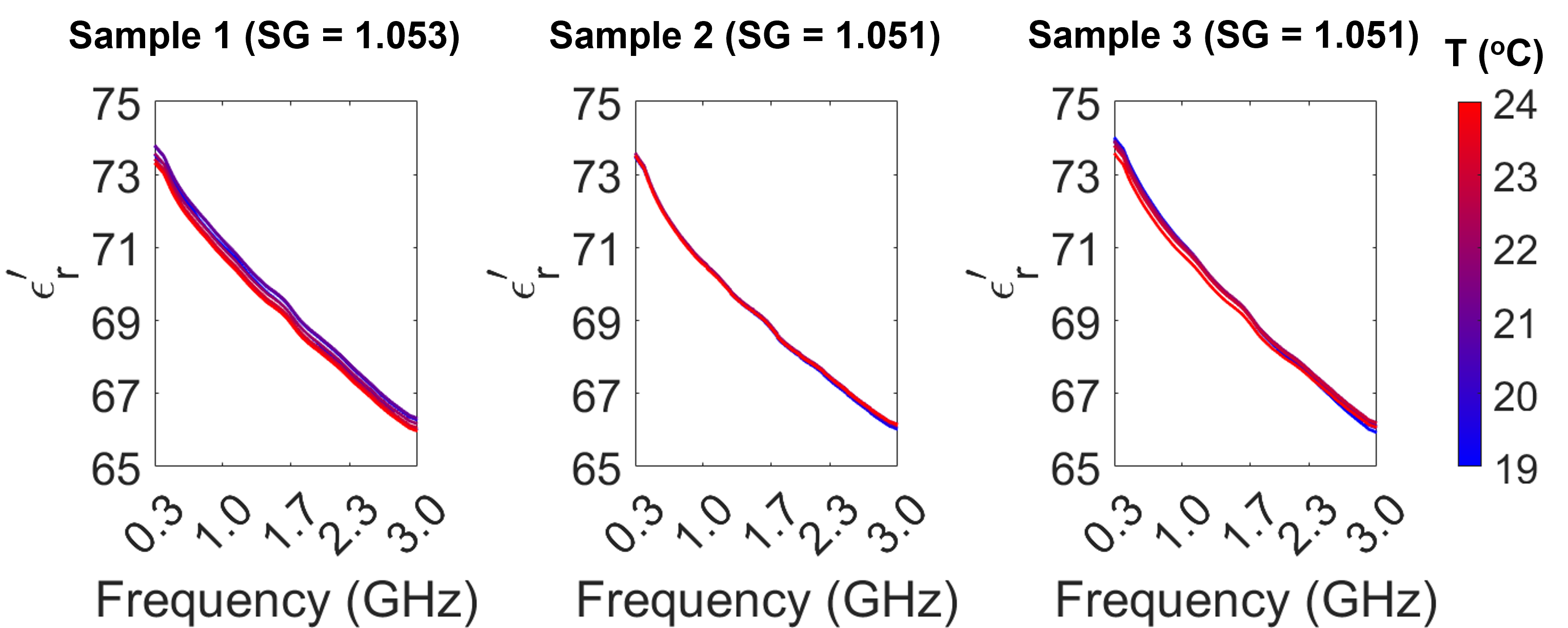}}
	\caption{Variation of the dielectric constant with temperature (19–24 °C) for three potato samples over the frequency range of 0.3–3.0 GHz. The transition of the line color from blue to red represents increasing temperature.}
	\label{fig:temp}
\end{figure}

To evaluate the influence of temperature variation on the dielectric constant of potato tubers, we measured the dielectric constant of three fresh yellow potatoes at temperatures ranging from 19 °C to 24 °C, recorded at 1 °C intervals. The three potatoes had an SG of around 1.05 and were prepared following the same process described in Section II.C. The results are shown in Fig. \ref{fig:temp}. The changes in dielectric constant across 19–24°C are small for all three samples across the frequency range. The maximum dielectric constant changes $\Delta\epsilon_r'$ are less than 0.49, 0.20, and 0.45 for the three samples. When applying the developed SG estimation model, the maximum estimation error resulted from the dielectric changes is less than \(2.27\times10^{-3}\) across all frequency points, which is very small and within the acceptable range. These results demonstrate that small temperature fluctuations do not significantly affect the dielectric constant of potato tubers, and therefore, our model is robust against such variations.\par

\subsection{Frequency Selection}

Potato tubers are primarily composed of water (73\%–77\%) and starch (18\%–20\%)~\cite{Starbook}. The dielectric response of such biological tissues can be modeled using a solid–water mixture model~\cite{Dbook1, Dbook2}. In this model, the overall dielectric properties are determined by the volume fraction of dry matter (mostly starch) and water as well as their respective dielectric characteristics. The volume fraction of dry matter and water can be represented by SG. The dielectric constant of starch is around 6.5 at 0.3 GHz, and only slightly decrease with increasing frequency~\cite{rstarch, rstarch2}. In contrast, the dielectric constant of water is as high as about 80 at low frequencies, but it decreases rapidly with increasing frequency due to dielectric relaxation.\par

In our experiments, we observed that the correlation between dielectric constant and SG weakens in the low-frequency range. This could be because, at low frequencies (below 300 MHz), the dielectric response of potatoes is significantly influenced by ionic conduction caused by dissolved ions~\cite{rstarch, rions}. This effect masks the dielectric variations associated with the starch-water ratio, making it difficult to separately analyze the relationship between the starch-water ratio and dielectric properties.\par

As frequency increases, the dielectric constant of water decreases due to dielectric relaxation~\cite{rwater1, rwater2, rwater3}, reducing its difference from the dielectric constant of starch ($\sim$5). Consequently, the sensitivity of the overall dielectric constant of potatoes to the starch-water ratio reduces with increasing frequency, which limits the accuracy of estimating SG based on dielectric constant at higher frequencies.\par

The 0.3-3.0 GHz range offers an optimal balance: it minimizes interference from ionic conduction while maintaining sensitivity of dielectric constant to SG.  Our experimental results confirm that within this frequency range, the correlation between dielectric constant and SG is strong and consistent across different potato types and growth stages. \par

\subsection{Future Applications}
SG is a critical parameter for evaluating the growth stage and quality of potato tubers, as well as for classifying potatoes for different food applications. The SG of potato tubers can vary by 0.007 to 0.030 over a 10-day period during the growing season ~\cite{SGrange2}. When the potatoes are harvested between 80 and 110 days after planting, the SG for potatoes can increase by more than 0.04~\cite{SGrange2}.\par 

The SG for a high-quality potato tuber varies depending on its intended use and differs among potato types. For example, potatoes used in chip and French fry production typically require a SG of 1.08 or higher~\cite{SG3}; potatoes intended for boiling or canning are preferred to have a lower SG, generally below 1.07~\cite{cook}. For the five types of potatoes we studied in this paper, red, yellow, and purple potatoes are generally lower in starch and better suited for boiling or baking. Their desired SG falls between 1.03 and 1.06. Russet and chipping potatoes are high-starch varieties primarily used for French fries or potato chips. The desired SG for russet potatoes ranges from 1.06 to 1.10, and for chipping potatoes ranges from 1.08 to 1.13.\par

In this study, we established a quantitative relationship between the dielectric constant and SG. Using the dielectric constant as input, the developed SG estimation model achieves high estimation accuracy. The model's performance allows for accurate tracking of SG changes during potato growth and supports classification for food processing purposes. In the future, integrating this model with advanced microwave sensing technologies could enable non-destructive, accurate SG measurements during both the growing season and the post-harvest storage season. For example, advanced ground-penetrating radar (GPR) technologies could be used to non-destructively measure the dielectric constant of potato tubers growing in the soil. Applying the developed model to these measurements would allow producers to non-destructively and efficiently monitor tubers' SG throughout the growing season. This information can then be used to optimize management practices, such as irrigation, fertilization, pest management, and harvesting, to reduce the environmental impact of potato production. Similarly, the model could be combined with microwave sensing techniques to non-invasively and rapidly assess potato SG in the food processing industry. Our future research will focus on developing practical non-destructive microwave sensing systems and applying the model to enable in-situ, non-destructive potato measurements.\par

\section{Conclusions}\label{sec5}
In this paper, we present a robust model for estimating the potato tubers' SG based on their dielectric constant at any frequency between 0.3 GHz and 3.0 GHz. The model was developed by fitting a linear regression model to the measured dielectric constant and SG values of 200 potatoes (five types, 40 samples each). To account for the dispersive nature of tuber tissues, the regression coefficients were modeled as a fourth-order polynomial function of frequency. The model was validated using measured data from 50 additional potatoes (five types, 10 samples each) that were not used in the model development and 50 yellow potatoes at different growing stages. Across the considered frequency range, the MAE and MAPE between the model-estimated and ground-truth SG values were consistently less than \(4.8\times10^{-3}\) and 0.45\%, respectively. The robust ability of the model to accurately estimate the SG of potato tubers supports the advancement of microwave sensing technologies for monitoring tuber growth in the soil, assessing potato quality, and classifying potatoes for different food applications. \par

\par

\end{document}